# Influence of misfit strain on the Goos-Hänchen shift upon reflection from a magnetic film on a non-magnetic substrate

Yu. S. Dadoenkova,[1,2,3,*] F. F. L. Bentivegna,[4] N. N. Dadoenkova,[1,3]
I. L. Lyubchanskii,[3] Y.P. Lee[5]

[1]Ulyanovsk State University, 432017 Ulyanovsk, Russian Federation
[2]Institute of Electronic and Information Systems, Novgorod State University, 173003 Veliky Novgorod, Russian Federation
[3]Donetsk Physical & Technical Institute of the National Academy of Sciences of Ukraine, 83114 Donetsk, Ukraine
[4]Lab-STICC (UMR CNRS 6285), ENIB, 29238 Brest Cedex 3, France
[5]Quantum Photonic Science Research Center (q-Psi) and Hanyang University, 04763 Seoul, South Korea
*Corresponding author: yulidad@gmail.com

The influence of the misfit strain on the lateral shift (Goos-Hänchen effect) experienced by a near-infrared electromagnetic wave upon reflection from the surface of a bilayer consisting of a magnetic, gyrotropic (i.e., whose permittivity tensor elements depend upon magnetization) yttrium-iron garnet film deposited on a non-magnetic gadolinium-gallium garnet substrate is investigated theoretically. In the geometry of the transverse magneto-optical Kerr effect, it is shown that the mechanical strain near the geometrical film/substrate interface can induce a significant lateral shift of the beam for incidence angles close to normal incidence, where no shift appears in the absence of strain. Our calculations demonstrate positive as well as negative values of the lateral shift, depending on the incident light polarization and on the film thickness. In contrast to that of the misfit strain, the influence of the magnetization of the gyrotropic film on the lateral shift is more noticeable for a TM- than for a TE-polarized wave.

## I. INTRODUCTION

The lateral shift of a reflected beam relative to its geometrical optics position was reported for the first time in 1947 by Goos and Hänchen in the case of the total internal reflection at the interface between two dielectric media [1, 2]. Since then, numerous theoretical and experimental works have been devoted to this phenomenon, expanding its scope to the reflection from virtually any type of surface and from any type of structure, including photonic crystals. Moreover, it can be extended to any kind of wave process and has been described for acoustic waves (Shoch effect) [3], spin waves [4-6], neutrons [7], and electrons [8]. Beyond their fundamental interest, Goos-Hänchen-like lateral shifts upon reflection or transmission of a light beam, as well as the related Imbert-Fedorov effect (*i.e.*, a beam shift in the direction perpendicular to the plane of incidence), have become relevant in technological domains, as illustrated by recent review papers [9, 10]. They have been shown to affect the propagation modes in optical waveguides or microcavities [11] designed for photonic applications, and have also been studied in many types of structures, including electro-optic [12-15] or magneto-optic materials [16-21], photonic crystals [22-24], superconducting multilayers [25, 26], plasmonic structures [27-29], graphene [30-32], at the interface between an ordinary dielectric and a topological insulator [33], as well as in metamaterials [34-38]. Recently the Goos-Hänchen effect has been studied for partially coherent light fields [39] and in a standing-wave-coupled electromagnetically-induced-transparency medium [40, 41]. The Goos–Hänchen effect has also been used successfully to design an optical waveguide switch [42], and in recent publications, applications to biosensing (*e.g.*, the measurement of E. Coli O157: H7 concentration [43]) or to the detection of chemical vapours [44] were reported. In magneto-optical materials the Goos–Hänchen shift can typically reach values up to several tens of light wavelengths [17]. Giant Goos–Hänchen shifts (equal to about one hundred light wavelengths) were observed in reflection from multilayered photonic crystals [24]. Thus the Goos–Hänchen shift ought to be taken into account for the design of integrated magneto-optical devices.

Obviously, the amplitude of the Goos-Hänchen shift depends upon such parameters as the angle of incidence, wavelength and state of polarization of the incident electromagnetic wave, but it also strongly depends on the nature and anisotropy of the media forming the interface, and, more generally, on that of all the media constituting the structure on which reflection takes place. The anisotropy of the materials can be inherent to their crystalline symmetries, but can also be externally controlled, such as in the case of magnetic media whose magnetization can be reversed. Moreover, the quality of the interfaces encountered by an electromagnetic wave can also be of importance. Indeed, it is well known that elastic strain takes place in the vicinity of interfaces in layered structures due to the crys-talline lattice misfit between the neighbouring materials [45, 46]. As a consequence, atomic alignment on each side of the geometric interfaces is altered, and the thickness of the resulting deformed layers can reach values up to a few hundred ångströms. Such crystalline deformations near interfaces are known to affect the optical properties, notably the reflectivity, of the multilayers *via* the photoelastic interaction [47]. A strong influence of epitaxial strain on the magneto-optical response of Co films was recently reported [48]. In our papers [49-51] we investigated the influence of misfit strain on optical second harmonic generation in terms of the nonlinear photoelastic tensor and the modulation of the quadratic nonlinear optical susceptibility of the medium. Similarly, we studied the effect of interfacial strain on the transmittivity of a dielectric photonic crystal [52].

The reflection of light from a magnetic material in which the magnetization is perpendicular to the plane of incidence reflection plane (transverse magneto-optic Kerr effect) is a powerful tool for the study of magnetic materials, surfaces and interfaces [53]. The Goos–Hänchen effect from a ferrite-ferrite interface in that configuration was for instance theoretically analyzed in Ref. [16], but misfit strain was not taken into account.

In this Paper, we propose a study of the lateral shift experienced by an electromagnetic wave upon reflection, in the transverse magneto-optical Kerr effect geometry, from the upper surface of a strained dielectric bilayer consisting of a magnetic film of yttrium-iron garnet (YIG) epitaxially grown on a non-magnetic substrate of gadolinium-gallium garnet (GGG) and surrounded by vacuum. The YIG/GGG bilayer has been a widely studied system for several decades. Garnets such as GGG and YIG exhibit a very good transparency in the near-infrared and infrared regimes of the electromagnetic spectrum. Moreover, these two crystals can easily be grown on top of each other, as their lattice constants only slightly differ — their difference leading, however, to misfit strain. Therefore the YIG/GGG bilayer has proved to be of particular interest for the design of magneto-optical structures and devices (magnetic photonic crystals [54, 55], Faraday rotators, magneto-optical switches, *etc.* [53]) in the near-infrared and infrared domains.

We limit ourselves to the Goos-Hänchen shift, but a similarly detailed study of the Imbert-Fedorov spatial and angular shifts in the same system and with the same approach poses no particular difficulty. Section II is devoted to the description of the geometry of the system and the photoelastic contribution of mechanical strain on the permittivity tensors of the materials on both sides of their geometrical interface. Section III describes the resolution of wave propagation in the inhomogeneous bilayer using the Green's functions techniques, and expresses the overall complex reflectivity of the system, whose phase leads to the determination of the lateral shift of the beam. In section IV, numerical simulations are performed for a near-infrared electromagnetic beam impinging on a bilayer consisting of a magnetic, gyrotropic yttrium-iron garnet film deposited on a non-magnetic gadolinium-gallium garnet substrate, and the influence of the mechanical strain near the geometrical film/substrate interface is studied and discussed, as well as that of the magnetization in the magnetic film. In Conclusion we summarize our results and an Appendix details the derivation of the Green's functions.

## II. DESCRIPTION OF THE SYSTEM

### A. Geometry

We consider a bilayer in vacuum consisting of a thin magnetic film (henceforth identified by index $\alpha = 1$) epitaxially grown on a non-magnetic substrate ($\alpha = 2$). The materials on either side of the interface have similar cubic crystal lattices, with lattice constants respectively denoted $a_1$ and $a_2$, yet the difference of these lattice constants leads to misfit strain and subsequent dislocations in the vicinity of their geometric interface. Figure 1 shows the geometry of the problem. The interface between the materials is parallel to the ($xy$) plane of a Cartesian system of coordinates. A plane wave of angular frequency $\omega$ impinges the surface of the magnetic film under oblique incidence angle $\theta$. The plane of incidence is ($xz$) and the incident and reflected waves can be decomposed into TE and TM components of electric field strengths $E_{\text{TE,TM}}^{(r,i)}$, where superscripts ($i$) and ($r$) correspond to the incident and reflected components, respectively. The magnetization vector in the magnetic film is directed along the $y$-axis, which corresponds to the transverse magneto-optical configuration. The thicknesses of the film and the substrate are denoted $D_1$ and $D_2$, respectively. Near the interface, a pseudomorphous region is established, in which the mechanical strain is homogeneous. The thicknesses of the pseudomorphous region on each side of that interface (called critical thicknesses) are denoted $h_c^{(1)}$ and $h_c^{(2)}$, respectively (see dotted lines in Fig. 1), so that its total thickness is $h_c^{(1)} + h_c^{(2)}$. If the thickness of the film is larger than $h_c^{(1)}$, misfit dislocations (indicated by symbols "⊥" in Fig. 1) appear above that region, where they tend to compensate for misfit strain and minimize the total mechanical strain energy.

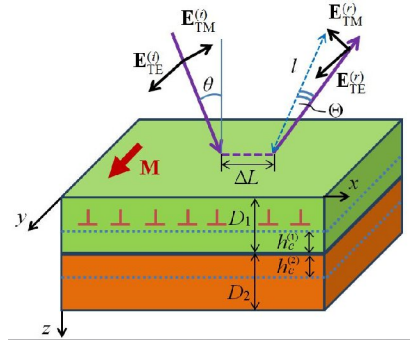

Fig. 1. Schematic of the Goos-Hänchen shift in the transverse magneto-optical geometry in the system under study. A magnetic film (thickness $D_1$) is epitaxially grown on a nonmagnetic substrate (thickness $D_2$). The thicknesses of the pseudomorphous regions on each side of their geometric interface are $h_c^{(1)}$ and $h_c^{(2)}$. The magnetization M = {0, $M_y$, 0} in the film is perpendicular to the plane of incidence ($xz$). The TE and TM components of the incident ($i$) and reflected ($r$) optical electric fields are denoted $E_{\text{TE,TM}}^{(r,i)}$. The spatial and angular Goos-Hänchen shifts in the plane of incidence upon reflection are $\Delta L$ and $\Theta$, respectively.

In the transverse magneto-optical configuration, the relative dielectric permittivity tensor assumes the following form:

$$\hat{\varepsilon}^{(1,0)} = \begin{pmatrix} \varepsilon_1 & 0 & -i\varepsilon' \\ 0 & \varepsilon_1 & 0 \\ i\varepsilon' & 0 & \varepsilon_1 \end{pmatrix}, \quad (1)$$

whereas in the non-magnetic substrate it simply can be written as

$$\varepsilon_{ij}^{(2,0)} = \varepsilon_2 \, \delta_{ij}, \quad (2)$$

where $\delta_{ij}$ is the Kronecker symbol.

It should be noted that if quadratic (in magnetization) magneto-optical contributions to the permittivity are neglected, the diagonal elements of $\hat{\varepsilon}^{(1,0)}$ do not depend on magnetization amplitude $M_y$, while its off-diagonal elements are linear in $M_y$ and can be written as $\varepsilon' = f'_e m_y$, where $f'_e$ is the gyroelectric coefficient of the magnetic film and $m_y = M_y / M_{sat}$ its reduced magnetization, $M_{sat}$ being its saturation magnetization [56]. Furthermore, we neglect a potential magnetic gyrotropy, *i.e.*, a dependence of the permeability of the magnetic film upon magnetization. In effect, in both materials, the permeability will be taken as that of the vacuum.

When a monochromatic Gaussian wavepacket impinges on the uppermost surface of the thin film, its reflection undergoes a non-negligible Goos-Hänchen lateral shift in the plane of incidence (in our case, along the *x*-axis) denoted $\Delta L$ in Fig. 1, as well as an angular shift Θ in the same plane. The lateral and angular shifts can be obtained with the stationary-phase approach proposed by Artmann [57]. Assuming an incident Gaussian beam of waist $w_0$ whose transverse field distribution along the *x*-axis writes:

$$E^{(i)}(x) = E_0 \exp\left[-\frac{x^2}{2w_0^2}\right], \quad (3)$$

the spatial profile $E^{(r)}(x)$ of the reflected beam is the inverse Fourier transform of the convolution product between the Fourier transform $\tilde{E}^{(i)}(K)$ of the incident field profile and the complex reflection function $R(k_x)$ of the system — where $k_x$ is the component of the wavevector along the *x*-axis, whose distribution of values stems from the angular divergence of the Gaussian beam:

$$E^{(r)}(x) = \frac{1}{2\pi}\int_{-\infty}^{\infty} \tilde{E}^{(i)}(K) R(K+k_c) \exp(iKx) dK,$$
$$\tilde{E}^{(i)}(K) = \int_{-\infty}^{\infty} E^{(i)}(x) \exp(-iKx) dx, \quad (4)$$

where $k_c$ is the central wavevector of the incident beam and $K = k_x - k_c$. The complex reflection function $R(k_x)$ can be written in the form:

$$R(k_x) = \exp\left[-i\left(\psi(k_x) + i \ln|R(k_x)|\right)\right], \quad (5)$$

where $\psi(k_x) = \arg[R(k_x)]$ is the phase difference between the reflected and incoming light waves. An approximate expressions for $E^{(r)}(x)$ can be obtained *via* a Taylor expansion of $\Phi(k_x) = \psi(k_x) + i \ln|R(k_x)|$ in the vicinity of $k_c$:

$$\Phi(k_x) \approx \Phi(k_c) + K\frac{\partial(\psi + i\ln|R|)}{\partial k_x} + \frac{K^2}{2}\frac{\partial^2(\psi + i\ln|R|)}{\partial k_x^2} + \ldots, (6)$$

where all derivatives are taken at $k_x = k_c$. Neglecting terms higher than the second order in the Taylor series, it can be shown that the lateral shift of the reflected wavepacket then simply writes:

$$\Delta L = -\frac{\partial \psi}{\partial k_x} - \frac{\partial \ln|R|}{\partial k_x}\frac{\partial^2 \psi}{\partial k_x^2}\left(w_0^2 + \frac{\partial^2 \ln|R|}{\partial k_x^2}\right)^{-1}, \quad (7)$$

while the angular shift can in first approximation be expressed as:

$$\Theta = -\left(\theta_0^2/2\right)\frac{\partial(\ln|R|)}{\partial k_x}. \quad (8)$$

Parameter $\theta_0 = \lambda_0/(\pi w_0)$ is the angular spread of the incident beam of wavelength $\lambda_0$ [58]. The Goos-Hänchen lateral shift $\Delta L$ is taken to be positive towards the positive direction of the *x*-axis, and the angular shift Θ is counted positive for a clockwise rotation with respect to the positive direction of the *y*-axis (in Fig. 1, both shifts are thus positive). It should be noted that the stationary phase approach is valid for incident beams with a sufficiently large beam waist, *i.e.*, with a narrow angular spectrum [57], in which case the reshaping of the beam upon reflection can be neglected. In this approximation, the reflected beam can then be considered Gaussian, with a sharp spectral distribution around $k_c$ [59]. Although Artmann derived the lateral shift for total internal reflection, it has been shown that this approach remains valid for the partial reflection of a well-collimated beam [60].

B. Influence of strain on the photoelastic properties

The value of the misfit strain due to the mismatch between the lattice constants $a_1$ and $a_2$ of the thin film and the substrate is proportional in each material to the corresponding misfit parameter $f^{(1)}$ (in the film) and $f^{(2)}$ (in the substrate), with

$$f^{(1)} = \frac{a_1 - a_2}{a_2}, \quad f^{(2)} = \frac{a_2 - a_1}{a_1}. \quad (9)$$

Consequently, a displacement of the atomic layers takes place with respect to their positions in the absence of strain, and is represented by the displacement vector u(r), where r = (*x*, *y*, *z*). This in turn induces a modification of the relative permittivity tensor of the material on each side of the geometric interface that can be written as:

$$\hat{\varepsilon}^{(\alpha)} = \hat{\varepsilon}^{(\alpha,0)} + \delta\hat{\varepsilon}^{(\alpha)}, \quad \alpha = (1, 2), \quad (10)$$

where $\hat{\varepsilon}^{(\alpha,0)}$ is the relative dielectric permittivity tensor of the strain-free material and the elements of the strain-induced contribution to its permittivity are, in a first-order approximation, given by

$$\delta\varepsilon_{ij}^{(\alpha)} = p_{ijkl}^{(\alpha)} u_{kl}^{(\alpha)}. \quad (11)$$

For the sake of simplicity, the superscript *α* denoting the material will sometimes be omitted in the following. In Eq. (11), as well as in the remainder of this Paper, summation over repeated indices is assumed, $p_{ijkl}^{(\alpha)}$ are the elements of the linear photo-elastic tensor of the material, and $u_{kl}^{(\alpha)}$ are the elements of its strain tensor. The latter are related to the displacement vector u in each material along:

$$u_{kl} = \frac{1}{2}\left(\frac{\partial u_k}{\partial r_l} + \frac{\partial u_l}{\partial r_k}\right), \quad (r_k, r_l) \in (x, y, z). \quad (12)$$

Both the materials constituting the film and the substrate are supposed to belong to the same cubic symmetry group ($O_h$), with the following non-zero photo-elastic tensor elements:

$$\begin{aligned}
p_{xxxx} &= p_{yyyy} = p_{zzzz} = p_{11}, \\
p_{xxyy} &= p_{yyxx} = p_{xxzz} = p_{zzxx} = p_{yyzz} = p_{zzyy} = p_{12}, \\
p_{xyxy} &= p_{yxyx} = p_{xyyx} = p_{yxxy} = p_{xzxz} = p_{zxzx} = p_{xzzx} \\
&= p_{zxxz} = p_{yzyz} = p_{zyzy} = p_{yzzy} = p_{zyyz} = p_{44}.
\end{aligned} \quad (13)$$

For the same reasons of cubic symmetry, the biaxial stress tensor describing the mechanical constraints parallel to the interface in each material can be written as:

$$\hat{\sigma} = \delta_{ij}\,\sigma_0(z). \quad (14)$$

Assuming an elastic deformation of the crystals, the strain tensor elements defined in Eq. (12) obey Hooke's law and have the following form:

$$u_{ij} = \frac{1}{E}\left[(1+\nu)\,\sigma_{ij} - \nu\,\mathrm{Tr}(\hat{\sigma})\,\delta_{ij}\right], \quad (15)$$

where $E$ and $\nu$ are the Young modulus and Poisson's ratio of the material, respectively. From Eqs. (14) and (15), the strain tensor elements thus read:

$$u_{xx} = u_{yy} \equiv u(z) = \frac{1-\nu}{E}\sigma_0(z),$$
$$u_{zz} = -\frac{2\nu}{E}\sigma_0(z) = -\frac{2\nu}{1-\nu}u(z) \quad (16)$$

Taking into account Eqs. (11), (13) and (16), it appears that only photo-elastic constants $p_{11}$ and $p_{12}$ actually play a role in the contribution of strain to the permittivity of the materials.

In the pseudomorphous layer strain is homogeneous, while in the regions of the bilayer outside that layer, strain is decreasing from the boundaries of the pseudomorphous region due to the appearance of misfit dislocations in the film. It has been reported earlier that strain tensor components can be estimated to decrease exponentially as the distance from the dislocation pileup plane increases [61], so that functions $u^{(\alpha)}(z)$ can be expressed as:

$$u^{(1)}(z) = \begin{cases} f^{(1)}\exp\left[\dfrac{z - D_1 + h_c^{(1)}}{h_c^{(1)}}\right], & 0 < z < D_1 - h_c^{(1)} \\ f^{(1)}, & D_1 - h_c^{(1)} < z < D_1 \end{cases}$$
$$u^{(2)}(z) = \begin{cases} f^{(1)}\exp\left[\dfrac{-z + D_1 + h_c^{(2)}}{h_c^{(2)}}\right], & D_1 + h_c^{(2)} < z < D_1 + D_2 \\ f^{(1)}, & D_1 < z < D_1 + h_c^{(2)} \end{cases} \quad (17)$$

Figure 2 shows the spatial variation of the strain in the bilayer as described by Eq. (17).

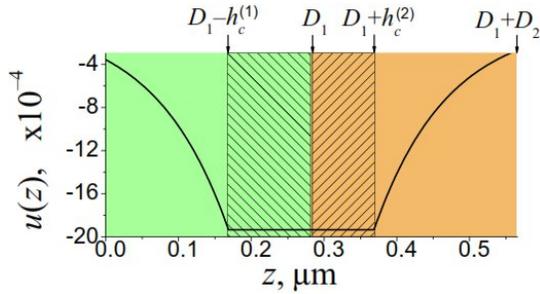

Fig. 2. Spatial variation of the misfit strain in the bilayer (see Table 1 for numerical values of the material properties). Shaded areas show the pseudomorphous layers in the film and the substrate.

In first approximation, the crystal symmetry in the strained regions can reasonably be assumed to remain unchanged, and in that case only diagonal elements of the strain-dependent contribution to the permittivity tensors of the film and the substrate are non-zero. Their expressions in each of the four regions of the bilayer defined in Eq. (17) are given by:

$$\delta\varepsilon_{xx}^{(\alpha)}(z) = \delta\varepsilon_{yy}^{(\alpha)}(z) = u^{(\alpha)}(z)\left[p_{11}^{(\alpha)} + p_{12}^{(\alpha)}\left(1 - \frac{2\nu^{(\alpha)}}{1-\nu^{(\alpha)}}\right)\right],$$
$$\delta\varepsilon_{zz}^{(\alpha)}(z) = u^{(\alpha)}(z)\left[2p_{12}^{(\alpha)} - p_{11}^{(\alpha)}\frac{2\nu^{(\alpha)}}{1-\nu^{(\alpha)}}\right]. \quad (18)$$

## III. ANALYTICAL RESOLUTION OF THE WAVE EQUATION

### A. Green's functions method

The aim of this Section is to show the technique used for the resolution of the wave equation in the inhomogeneous structure resulting from strain-dependent perturbations of the permittivity:

$$\mathbf{grad}(\mathrm{div}\,\boldsymbol{E}) - \Delta\boldsymbol{E} - k_0^2\,\hat{\varepsilon}^{(0)}\boldsymbol{E} = k_0^2\,\delta\hat{\varepsilon}(z)\boldsymbol{E}, \quad (19)$$

where the expressions of the elements of the perturbation tensor $\delta\hat{\varepsilon}(z)$ are given by Eq. (16). Note that this equation remains valid in a homogeneous medium such as vacuum, in which that tensor is simply zero. In inhomogeneous media where its elements depend upon $z$, solutions of Eq. (19) can be obtained using the Green's functions approach.

As a first stage, the solution of the homogeneous wave equation must be obtained using the specific form of the unstrained relative permittivity tensor $\hat{\varepsilon}^{(\alpha,0)}$ of each medium, defined by Eqs. (1)-(2).

Assuming solutions of the *homogeneous* wave equation to be plane waves, spatial Fourier transform applied to *inhomogeneous* Eq. (19) leads to the following linear system of differential equations in each medium, with $\alpha = (1, 2)$:

$$\hat{L}^{(\alpha)}(z)\begin{pmatrix}E_x^{(\alpha)}(z)\\E_y^{(\alpha)}(z)\\E_z^{(\alpha)}(z)\end{pmatrix} = k_0^2\,\delta\hat{\varepsilon}^{(\alpha)}(z)\begin{pmatrix}E_x^{(\alpha)}(z)\\E_y^{(\alpha)}(z)\\E_z^{(\alpha)}(z)\end{pmatrix}. \quad (20)$$

The propagation operator $\hat{L}^{(\alpha)}(z)$ in either medium is:

$$\hat{L}^{(\alpha)}(\partial_z) = \begin{pmatrix} -k_0^2\varepsilon_\alpha - \partial_z^2 & 0 & i(k_x\partial_z - k_0^2\varepsilon'_\alpha) \\ 0 & k_x^2 - k_0^2\varepsilon_\alpha - \partial_z^2 & 0 \\ i(k_x\partial_z + k_0^2\varepsilon'_\alpha) & 0 & k_x^2 - k_0^2\varepsilon_\alpha \end{pmatrix}, \quad (21)$$

where differential operators $\partial_z \equiv \dfrac{\partial}{\partial z}$ and $\partial_z^2 \equiv \dfrac{\partial^2}{\partial z^2}$ are introduced. Note that $\varepsilon'_2 = 0$ in the nonmagnetic substrate.

The second stage of the procedure consists in introducing in each region of the structure a Green's function, *i.e.*, a perturbative solution of the inhomogeneous wave equation for a point source term localized at position $z'$. Such functions $G_{ij}^{(\alpha)}(z-z')$, $(i,j) \in (x,y,z)$ lead then to the expression of the general solution of Eq. (19), with [62]:

$$E_i(z) = E_i^{(0)}(z) + k_0^2\int_0^{D_1}G_{ik}^{(1)}(z-z')\,\delta\varepsilon_{kj}^{(1)}(z')\,E_j^{(1,0)}(z')\,dz' +$$
$$+ k_0^2\int_{D_1}^{D_1+D_2}G_{ik}^{(2)}(z-z')\,\delta\varepsilon_{kj}^{(2)}(z')\,E_j^{(2,0)}(z')\,dz'. \quad (22)$$

In this expression, fields $\boldsymbol{E}^{(\alpha,0)}$ are solutions of the homogeneous wave equation, and Green's functions are themselves solutions of the following equation:

$$\hat{L}_{ik}^{(\alpha)}(\partial_z) G_{kj}^{(\alpha)}(z-z') = \delta_{ij}\,\delta(z-z'), \quad (23)$$

where $\delta(z-z')$ is the Dirac function of variable $z$ denoting the presence of a point source at $z = z'$.

The detailed procedure for solving Eq. (23) is presented in the Appendix.

## B. Reflection coefficients for TE- and TM-polarized incoming light

From that point on, expressions of the optical electric field in all regions can be obtained using Eq. (22) and the Green's functions obtained in Appendix.

For the specific purpose of this work, we then need first to establish the reflection matrix of the vacuum/magnetic film interface at $z = 0$. Denoting $\boldsymbol{E}^{(i)}$ the field of the incoming electromagnetic wave in the vacuum, it is useful to express the components of its reflected counterpart as follows:

$$E_i^{(r)} = E_i^{(0,r)} + E_i^{(str,r)} = R_{ik}^{(0)} E_k^{(i)} + R_{ik}^{(str)} E_k^{(i)}, \quad (24)$$

*i.e.*, to separate the contribution of the unstrained bilayer from the perturbation induced by strain.

The reflection matrix at the interface writes then simply:

$$R_{ik} = R_{ik}^{(0)} + R_{ik}^{(str)}. \quad (25)$$

The determination of the reflection matrix of the unstrained structure is readily obtained, for instance following the classical techniques described by Yeh [63]. The strain-dependent reflection matrix is then deduced from the sole perturbative contribution to the reflected field in vacuum included in the following expression:

$$E_i^{(str)}(z) = k_0^2 \int_0^{D_1} G_{ik}^{(V)}(z-z')\,\delta\varepsilon_{kj}^{(1)}(z')\,E_j^{(1,0)}(z')\,dz' \\ + k_0^2 \int_{D_1}^{D_1+D_2} G_{ik}^{(V)}(z-z')\,\delta\varepsilon_{kj}^{(2)}(z')\,E_j^{(2,0)}(z')\,dz', \quad (26)$$

estimated at $z = 0$. In Eq. (26), $G_{ij}^{(V)}$ in each integral is the Green's function in vacuum for $z < 0$ deduced in Appendix from Eq. (A12a), with amplitudes $B_{ij}^{(V)}$ obtained using Eqs. (A9)–(A11) for a point source in the film and in the substrate, respectively.

Due to the symmetry of the crystals and the additional anisotropy imposed by the transverse magnetization in the magnetic film, the TE and TM states of polarization are eigenmodes of the structure, whose non-zero components of their electric and magnetic fields are ($E_y$, $H_x$, $H_z$) and ($E_x$, $E_z$, $H_y$), respectively. Thus, for any incoming state of polarization, the reflection matrix can be reduced to a (2×2) diagonal matrix that connects the amplitudes of the TE and TM components $E_{\text{TE,TM}}^{(r)}$ of the reflected field to those of the incident field $E_{\text{TE,TM}}^{(i)}$ as:

$$\begin{pmatrix} E_{\text{TE}}^{(r)} \\ E_{\text{TM}}^{(r)} \end{pmatrix} = \begin{pmatrix} R_{\text{TE}} & 0 \\ 0 & R_{\text{TM}} \end{pmatrix} \begin{pmatrix} E_{\text{TE}}^{(i)} \\ E_{\text{TM}}^{(i)} \end{pmatrix}, \quad (27)$$

where the complex reflection coefficients $R_{\text{TE}}$ and $R_{\text{TM}}$ of the TE and TM components of the electromagnetic radiation are directly related to the elements of the reflection matrix [Eq. (25)], with $R_{\text{TE}} = R_{yy}$ and $R_{\text{TM}} = \sqrt{R_{xx}^2 + R_{zz}^2}$.

## IV. NUMERICAL SIMULATIONS AND DISCUSSION

In this Section, we illustrate the theoretical calculations detailed above in the case of a magnetic layer of YIG ($Y_3Fe_5O_{12}$) grown on a substrate of GGG ($Gd_3Ga_5O_{12}$). Table 1 presents the dimensions of the structures and the material parameters used for these simulations.

Table 1. Physical data used for calculations

| | |
|---|---|
| Wavelength (nm) | $\lambda_0 = 1150$ |
| Lattice constants (Å) [64] | $a_1 = 12.376$ (YIG)    $a_2 = 12.40$ (GGG) |
| YIG relative permittivity tensor elements (at $\lambda_0$) [56] | $\varepsilon_1 = 4.5796$    $\varepsilon' = f'_e\,m_y = -2.47\times10^{-4}\,m_y$ |
| GGG relative permittivity (at $\lambda_0$) [56] | $\varepsilon_2 = 3.7636$ |
| Poisson's ratios [65] | $\nu^{(1)} = 0.2857$ (YIG)    $\nu^{(2)} = 0.28$ (GGG) |
| Non-zero linear photoelastic tensor elements for YIG [65] | $p_{11} = 0.025$,    $p_{12} = 0.073$,    $p_{44} = 0.041$ |
| Non-zero linear photoelastic tensor elements for GGG [64] | $p_{11} = -0.086$,    $p_{12} = -0.027$,    $p_{44} = 0.078$ |

Estimating the critical thickness $h_c^{(\alpha)}$ of the pseudomorphous layer in each material can be performed on the basis of mechanical arguments in the crystalline lattice [44]. The value of $h_c^{(\alpha)}$ in either material must satisfy:

$$h_c^{(\alpha)} = \frac{b^{(\alpha)}\left(1-\nu^{(\alpha)}\cos^2\beta\right)}{8\pi\left(1+\nu^{(\alpha)}\right) f^{(\alpha)}}\,\ln\frac{2h_c^{(\alpha)}}{b^{(\alpha)}}, \quad (28)$$

where $b^{(\alpha)}$ is the modulus of the Burgers vector of the lattice (for edge dislocations, it is equal to the lattice constant of the unstrained crystal) and $\beta = \pi/2$ for the edge dislocations that appear in YIG and GGG crystals. Using material parameters gathered in Table 1, Eq. (28) leads to the following values for the critical thickness of the pseudomorphous layer in each material, namely, $h_c^{(1)} \approx h_c^{(2)} = 0.1\,\mu\text{m}$.

For the calculations presented below for both TE- and TM-polarized incident waves, film and substrate thicknesses were taken as integer multiples of $d_1 = 0.269\,\mu\text{m}$ and $d_2 = 0.296\,\mu\text{m}$, respectively. These values correspond to both the film and the substrate being of equal half-wave optical thickness at normal incidence, *i.e.*, $d_1\sqrt{\varepsilon_1} = d_2\sqrt{\varepsilon_2} = \lambda_0/2$. This choice was made in order to emphasize the effect of strain on reflection, in conditions where the latter is near zero — *i.e.*, in the vicinity of normal incidence, as well as around the Brewster incidence angle for a TM-polarized wave. Note that if the bilayer were to form the unit cell of a photonic crystal, such a choice of layer thicknesses would correspond to the center of the photonic bandgap at normal incidence.

Unless otherwise specified, all calculations are performed for magnetization in the magnetic film saturated in the direction parallel to the $y$-axis, *i.e.*, for $m_y = +1$.

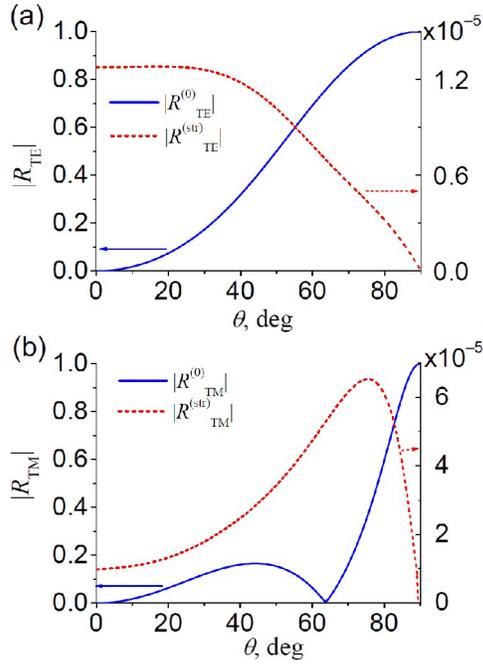

Fig. 3. Modulus of the reflection coefficient of the bilayer as a function of the angle of incidence $\theta$ for a TE-polarized (a) and a TM-polarized (b) incident wave. The solid blue line represents $R_{\text{TE,TM}}^{(0)}$ without taking film/substrate interfacial strain into account. The dashed red line shows the modulus of the correction $R_{\text{TE,TM}}^{(str)}$ to that reflection coefficient due to strain.

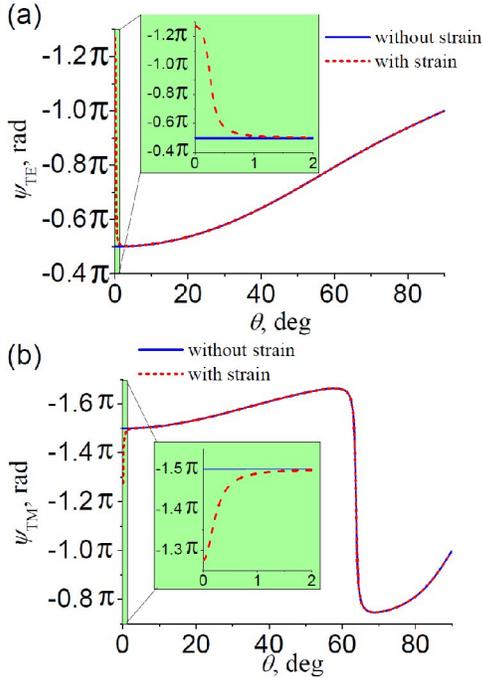

Fig. 4. Phase shift $\psi$ between the incoming and reflected waves as a function of the angle of incidence $\theta$ for a TE-polarized (a) and a TM-polarized (b) incident wave. The solid blue line represents the phase shift without taking film/substrate interfacial strain into account. The dashed red line shows its value when strain is accounted for.

Figure 3 shows the modulus of the reflection coefficient of the bilayer as a function of the angle of incidence $\theta$ for a TE-polarized (a) and a TM-polarized (b) incident wave, as well the modulus of the strain-dependent correction to that coefficient. As expected, $R_{\text{TE}}^{(0)}$ is zero at normal incidence. It should be noted that the reflection coefficient $R_{\text{TM}}^{(0)}$, on the other hand, is not strictly zero at normal incidence, due to the magnetic anisotropy that introduces gyrotropic terms in the dielectric permittivity tensor [Eq. (18)].

For the same reason, $R_{\text{TM}}^{(0)}$ only approaches zero at $\theta \approx 63.75°$, which is not the Brewster angle as commonly defined (*i.e.*, leading to a disappearance of the reflected TM wave [47]), but a pseudo-Brewster angle for which the reflection coefficient reaches a near-zero minimum. Due to the gyrotropy of the magnetic film, the reflection coefficient is complex, and its real and imaginary parts reach zero at slightly different incidence angles. The correction to the reflection coefficient due to strain $R_{\text{TE,TM}}^{(str)}$ is of the same order of magnitude for both TE and TM modes, and about four orders of magnitude smaller than the reflection coefficient of the strain-free structure. It was checked that conservation of energy is satisfied for all angles of incidence, including at grazing incidence where the strain-dependent corrections reach zero and the strain-free reflection coefficients are unity.

As mentioned earlier, in the stationary-phase approach, the calculation of the lateral shift of a Gaussian reflected beam requires the knowledge of the phase shift between incident and reflected wave for all angles of incidence within the divergent beam [see Eq. (1)]. This phase shift $\psi$ is none other than the argument of the reflection coefficient. Figure 4 shows its angular dependence for a TE-polarized (a) and a TM-polarized (b) incident wave, with and without film/substrate interfacial strain.

The impact of strain is only significant in the vicinity of the normal incidence, where the phase shift exhibits a steep variation from $-1.25\pi$ to $-0.5\pi$ for a TE-polarized wave, and from $-1.25\pi$ to $-1.5\pi$ for a TM-polarized wave, as one can see from the insets in Fig. 4. It can thus be expected that the strain-dependent contribution to the lateral shift of the reflected beam will also assume significant values near normal incidence.

Indeed, as Fig. 5 shows, the strain-dependent correction to that shift is only noticeable for small values of the incidence angle. However, in that region, this contribution reaches large values. The strain-induced lateral shift exceeds 100 wavelengths of the incident light to the left with respect to the *x*-axis (one speaks of a negative shift) for a TE-polarized wave, whereas it reaches about 18 wavelengths to the right (positive shift) for a TM-polarized wave, as can be seen on the left insets in Fig. 5. In the latter case, an even larger lateral shift (about 55 wavelengths to the left) occurs at the pseudo-Brewster angle. The right inset in Fig. 5(b) shows that the angular position of the extremal value of the lateral shift is slightly different when strain is taken into account and when it is not — this corresponds to a shift of the pseudo-Brewster angle of about 0.02°.

Since the lateral shift depends on the phase difference between the incoming and the overall reflected waves, its sign (positive or negative) is governed by the phase-dependent interference of the waves reflected from the interfaces separating the different layers of the system [66]. For given values of the layer thicknesses, it can thus switch sign when the angle of incidence varies.

For both states of polarization, the lateral shift increases — whether strain is taken into account or not — when the incoming beam tends toward grazing incidence. This can be seen as the analog of the behavior of the Goos-Hänchen effect, which has been shown to be maximal near the angu-lar condition of total internal reflection at the interface separating a high-index medium from a low-index one. In our case, the increase toward unity of the reflectivity of the structure near grazing incidence can be seen as the equivalent of that behavior.

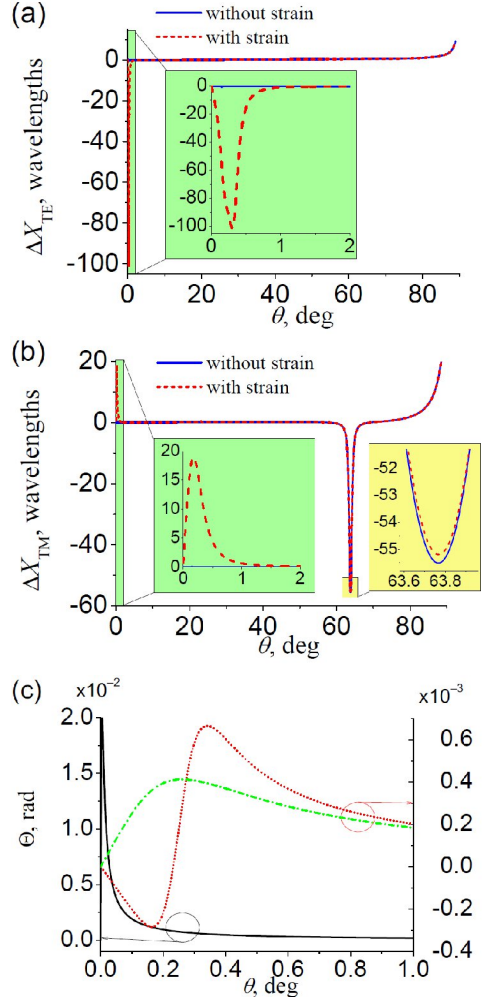

Fig. 5. Reduced Goos-Hänchen shift $\Delta X_{TE,TM} = \Delta L_{TE,TM} / \lambda_0$ as a function of the angle of incidence $\theta$ for a TE-polarized (a) and a TM-polarized (b) incident wave. The solid blue line represents the lateral shift without taking film/substrate interfacial strain into account. The dashed red line shows its value when strain is accounted for. (c) Angular Goos-Hänchen shift $\Theta$ near normal incidence without strain (solid black line for both TE- and TM-polarized light) and with strain for TE- (red dotted line) and TM- (green dash-dotted line) waves for a beam waist $w_0 = 100$ μm.

The angular Goos-Hänchen shift $\Theta$ in the plane of incidence is shown in Fig. 5(c) for an incident beam waist $w_0 = 100$ μm. In the absence of strain $\Theta$ is almost identical for TE and TM polarizations (black solid line) and is only significant in the vicinity of normal incidence. The presence of strain leads to an overall reduction of $\Theta$, and the additional anisotropy and inhomogeneity it introduces result in $\Theta$ alternating between negative and positive values for a TE-polarized beam (red dotted line), whereas it remains positive for a TM-polarized wave (green dash-dotted line). The total beam displacement observed at a distance $l$ from the geometrical point of incidence on the upper surface is $\Delta L + l \tan\Theta \approx \Delta L + l\Theta$ (see Fig. 1). The results presented in Fig. 5 show that the angular contribution to this displacement only becomes comparable to $\Delta L$ (at incidence angles where the latter is maximal) for $l > 10$ cm. In the following, we will thus restrict our analysis to the spatial Goos-Hänchen shift only.

The magnetic anisotropy in the magnetic film is also expected to have an impact on the optical response of the structure. The results in Figs. 3–5 were all obtained with the magnetic film magnetized at saturation in the positive direction of the $y$-axis ($m_y = +1$). Figure 6 presents the angular dependence of the variation of the reduced lateral shift $\Delta X_{TM}^{M+} - \Delta X_{TM}^{M-}$ upon complete reversal of the magnetization in the film, i.e., when magnetization is switched, so that $m_y = -1$ and off-diagonal components of the permittivity tensor Eq. (1) change sign, in the case of a TM-polarized incident wave. No variation of the lateral shift takes place for a TE-polarized wave, because the dependence of the permeability tensor of the magnetic film upon magnetization was neglected (in other words, we considered the film to be gyrotropic, but not bigyrotropic). This choice is justified by the fact that the off-diagonal elements of the permeability tensor of a bigyrotropic medium are at least one order of magnitude smaller than those of its permittivity tensor, hence are often neglected.

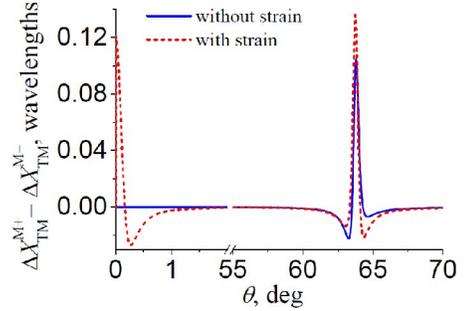

Fig. 6. Difference between the reduced Goos-Hänchen shifts $\Delta X_{TM}^{M+}$ and $\Delta X_{TM}^{M-}$ obtained for a TM-polarized incident wave with saturation magnetization in the magnetic film parallel and anti-parallel to the $y$-axis, respectively. The solid blue line represents the difference without taking substrate/film interfacial strain into account. The dashed red line shows its value when strain is accounted for.

Not surprisingly, the variation of the shift is virtually zero at angles of incidence for which the lateral shift was already negligible. Thus this variation is only noticeable around the pseudo-Brewster angle for the unstrained structure, and around that angle and near normal incidence when strain is taken into account. In all cases, the variation of $\Delta L$ upon magnetization reversal is about two orders of magnitude smaller than the shift before reversal, i.e., magnetization in the magnetic field plays a clearly lesser role in obtaining a lateral shift of the reflected beam than strain.

Finally, the dependance of the Goos-Hänchen shift upon the thicknesses of the film and of the substrate can be studied. Figures 7(a) and 7(b) show the evolution of the reduced shifts for TE and TM polarization, respectively, in the vicinity of their extrema near normal incidence, for increasing values of film thickness. Similarly, Figure 8 represents the same evolution for increasing values of the substrate thickness. Note that in both cases, film and substrate thicknesses increase as integer multiples of the nominal values given above (i.e., $D_i = N_i \times d_i$, $N_i \in \mathbb{N}$, $i \in \{1,2\}$, with $d_1 = 0.269$ μm and $d_2 = 0.296$ μm), thus preserving the half-wave condition for the optical thickness of the layers. Furthermore, in each case the substrate thickness is kept comparable to the film thickness, in order to remain in agreement with the theoretical description developed in this Paper.

As can be observed from Figs. 7, the dependence of the lateral shift upon film thickness $D_1$ is not monotonous for either TE or TM polarization. Indeed, for increasing values of $D_1$ the lateral shift of a TE-polarized wave changes sign, from negative ($\Delta X_{TE} = -130\, \lambda_0$ for $D_1 = 4 \times d_1$) to positive ($\Delta X_{TE} = 320\, \lambda_0$ for $D_1 = 5 \times d_1$) values [see green and cyan profiles in Fig. 7(a)]. Note that for still larger values of the film thickness, i.e., for $D_1 > 5 \times d_1$, the peak value of $\Delta X_{TE}$ keeps increasing due to the manifestation of the reflection coefficient minimum near normal incidence. In comparison, for a TM-polarized wave, $\Delta X_{TM}$ first exhibits an increase with the film thickness, then its peak starts to decay for $D_1 = 5 \times d_1$, and becomes negligible for $D_1 > 9 \times d_1$. The angular position of the extremum of $\Delta X_{TM}$ with respect to the film thickness monotonously drifts towards larger incidence angles for a TM-polarized wave. For a TE-polarized wave, on the other hand, the anisotropy is such that the position of the extremum of $\Delta X_{TE}$ exhibits a non-monotonous evolution with film thickness (even the number of extrema varies, as can be seen in Fig. 7(a), where $\Delta X_{TE}$ exhibits two negative maxima for some film thicknesses, e.g., for $D_1 = d_1$ and $D_1 = 4 \times d_1$).

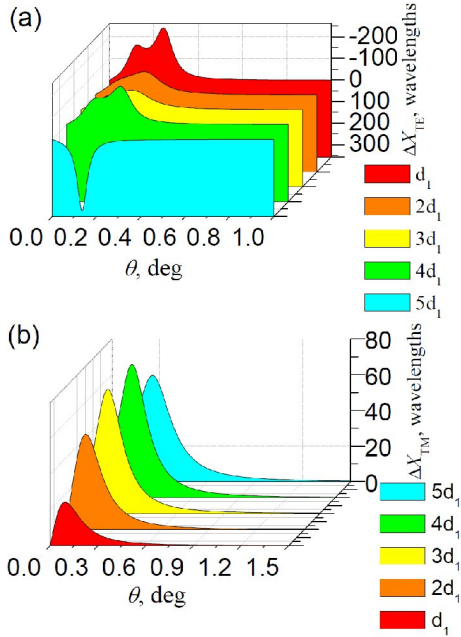

Fig. 7. Reduced Goos-Hänchen shifts (a) $\Delta X_{TE}$ and (b) $\Delta X_{TM}$ as functions of the angle of incidence $\theta$ for increasing values of the film thickness in the vicinity of the normal incidence. Here the nominal film thickness is $d_1 = 0.269\ \mu m$, and the substrate thickness is fixed at $D_2 = 4 \times 0.296\ \mu m$. Note the reversal of the vertical scale between (a) and (b).

Contrary to the dependence of the amplitude and position of the Goos-Hänchen shift peak near normal incidence upon the film thickness, their dependence upon the substrate thickness thickness $D_2$ is monotonous for both states of incoming polarization (Fig. 8). Indeed, for both TE- and TM-polarized light, the amplitude of the peak increases with $D_2$ and its position continuously shifts toward slightly larger angles of incidence. For certain values of the substrate thickness (for $D_2 > 3 \times d_2$), $\Delta X_{TE}$ again exhibits two negative maxima [see Fig. 8(a)].

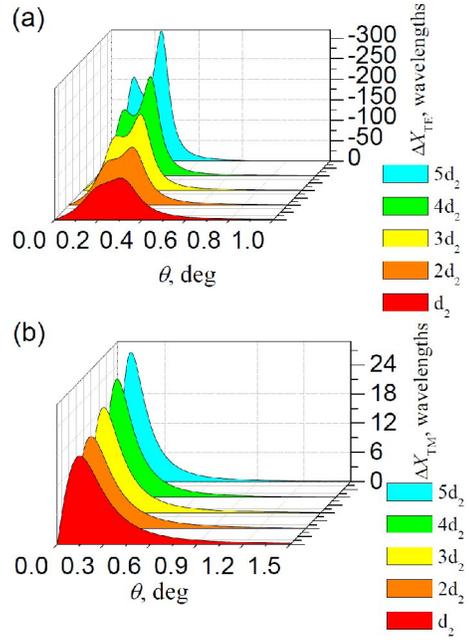

Fig. 8. Reduced Goos-Hänchen shifts (a) $\Delta X_{TE}$ and (b) $\Delta X_{TM}$ as functions of the angle of incidence $\theta$ for increasing values of the substrate thickness in the vicinity of the normal incidence. Here the magnetic film thickness is fixed at $D_1 = 0.269\ \mu m$ and the nominal substrate thickness is $d_2 = 0.296\ \mu m$. Note the reversal of the vertical scale between (a) and (b).

It must be emphasized that there would not be any peak of the Goos-Hänchen shift near normal incidence if it were not for strain in the epitaxially-grown film. The complex dependence of both the amplitude and the position of the lateral shift extremum upon incidence angle and light polarization can be assigned to the anisotropic influence of strain on the photo-elastic properties of both materials and, as a result, on the phase-dependent interference pattern of the multiple waves reflected from the various interfaces of the system. This influence is in no way straightforward, as the Goos-Hänchen shift depends on the variation of the phase-shift of the overall reflected wave with respect to the $x$-component of the wavevector — for an incoming divergent beam, the distribution of $k_x$ values will change upon reflection on each interface separating the optically and/or magnetically anisotropic layers of the system.

## V. CONCLUSIONS

We theoretically studied the influence of the misfit strain on the Goos-Hänchen (lateral) shift experienced by a near-infrared electromagnetic wave upon reflection from the surface of a dielectric bilayer consisting of a magnetic YIG film epitaxially grown on a non-magnetic GGG substrate. We showed that the mechanical strain near the geometrical film/substrate interface can be significant, and that the lateral shift can reach values from a few tens to hundreds of incident light wavelengths at incidence angles close to the normal incidence, as well as at the pseudo-Brewster angle. We demonstrated that the largest Goos-Hänchen shift, observed at small incidence angles, where the half-wave condition for both layers is nearly satisfied, is entirely induced by strain. It should be noted that if the half-wave condition is satisfied for an oblique incidence angle away from the normal incidence, the reflection coefficient turns to zero in the absence of strain, which results in a large Goos-Hänchen shift near this incidence angle. In this case, unlike what is obtained at the pseudo-Brewster angle, the impact of strain is large and strain-induced contributions to the reflection coefficient and its phase result in a broadening of the peak of the Goos-Hänchen shift. Our calculations also demonstrated that the lateral shift can be positive as well as negative, depending on the polarization of the incident light and on the thickness of the magnetic film. The influence of the magnetization of the YIG film (in the transverse magneto-optical configuration) on the Goos-Hänchen effect was also considered. For a TM-polarized incident wave, the variation of the lateral shift upon magnetization reversal was found to be only noticeable around the pseudo-Brewster angle for the unstrained structure, and around that angle and near normal incidence when strain is taken into account. No noticeable dependence of the lateral shift upon magnetization was observed for a TE-polarized wave.

# APPENDIX

## A. Derivation of the Green's functions

In order to obtain the Green's functions which provide analytical solutions for Eqs. (24), we transform them into a system of equations for which solutions are known, namely:

$$\left(\frac{d^2}{dz^2}+a^2\right)\xi(z)=2ia\delta(z),\quad \xi(z)=e^{ia|z|},$$
$$\left(\frac{d^2}{dz^2}+a^2\right)\zeta(z)=2\frac{d}{dz}\delta(z),\ \zeta(z)=e^{ia|z|}\,\text{sgn}(z).$$
(A1)

This can be achieved by noting that in both the anisotropic magnetic film and the isotropic non-magnetic substrate, the set of Eqs. (24) can be shown to be equivalent, through linear combinations, to:

$$\hat{U}_{\text{TM}}^{(\alpha)}(\partial_z)\begin{pmatrix} G_{xx}^{(\alpha)} & G_{xz}^{(\alpha)} \\ G_{zx}^{(\alpha)} & G_{zz}^{(\alpha)} \end{pmatrix} = \begin{pmatrix} L_{xx}^{(\alpha)} & -L_{xz}^{(\alpha)} \\ -L_{zx}^{(\alpha)} & L_{zz}^{(\alpha)} \end{pmatrix} \delta(z-z') \quad \text{(A2a)}$$

for TM polarization, and

$$\hat{U}_{\text{TE}}^{(\alpha)}(\partial_z)\, G_{yy}^{(\alpha)} = L_{yy}^{(\alpha)}\,\delta(z-z') \quad \text{(A2b)}$$

for TE polarization, where the spatial variation of Green's functions $G_{ij}^{(\alpha)}(z-z')$ has been omitted for the sake of simplicity.

Operators $\hat{U}_{\text{TE, TM}}^{(\alpha)}(\partial_z)$ in Eqs. (A2) are defined as

$$\hat{U}_{\text{TE}}^{(1)}(\partial_z) = -\partial_z^2 - k_{\text{TE}}^2,$$
$$\hat{U}_{\text{TM}}^{(1)}(\partial_z) = k_0^2 \varepsilon_1 \left[\partial_z^2 + k_{\text{TM}}^2\right], \quad \text{(A3)}$$
$$\hat{U}_{\text{TM}}^{(2)}(\partial_z) = \hat{U}_{\text{TE}}^{(2)}(\partial_z) = k_0^2 \varepsilon_2 \left[\partial_z^2 + k_{2z}^2\right],$$

where wavevectors $k_{\text{TE}}$ and $k_{\text{TM}}$ in the magnetic film, and the $z$-axis component of the wavevector in the non-magnetic substrate are given by

$$k_{\text{TE}}^2 = k_0^2\,\varepsilon_1 - k_x^2,\quad k_{\text{TM}}^2 = k_0^2\,\frac{\varepsilon_1^2-\varepsilon'^{\,2}}{\varepsilon_1}-k_x^2,\quad \text{(A4)}$$

$$k_{2z}^2 = k_0^2\,\varepsilon_2 - k_x^2. \quad \text{(A5)}$$

The set of differential equations Eqs. (A2), with the form assumed by operators $\hat{U}_{\text{TE, TM}}^{(\alpha)}(\partial_z)$ in Eqs. (A3), is similar to Eqs. (A1), which makes it then possible to obtain, in each region of the bilayer, for each localization $z'$ of the point source and for each state of polarization, the following analytical expressions that take into account general plane-wave solutions of the homogeneous equations and particular solutions of the inhomogeneous problem:

1) For $0 < (z,z') < D_1$:
   a) $0 < z < z'$: with $i=(x,z),\ k=(x,y,z)$,
   
   $$G_{ik}^{(1)}(z-z') = A_{ik}^{(1)} e^{ik_{\text{TM}}z} + B_{ik}^{(1)} e^{-ik_{\text{TM}}z}, \quad \text{(A6a)}$$
   
   $$G_{yk}^{(1)}(z-z') = A_{yk}^{(1)} e^{ik_{\text{TE}}z} + B_{yk}^{(1)} e^{-ik_{\text{TE}}z}. \quad \text{(A6b)}$$
   
   b) $z' < z < D_1$: with $i=(x,z),\ k=(x,y,z)$,
   
   $$\begin{aligned}G_{ik}^{(1)}(z-z') &= \left[A_{ik}^{(1)}+\lambda_{ik}^{(1)+}e^{-ik_{\text{TM}}z'}\right]e^{ik_{\text{TM}}z}\\&+\left[B_{ik}^{(1)}-\lambda_{ik}^{(1)-}e^{ik_{\text{TM}}z'}\right]e^{-ik_{\text{TM}}z},\end{aligned} \quad \text{(A6c)}$$
   
   $$\begin{aligned}G_{yk}^{(1)}(z-z') &= \left[A_{yk}^{(1)}+\lambda_{yk}^{(1)+}e^{-ik_{\text{TE}}z'}\right]e^{ik_{\text{TM}}z}\\&+\left[B_{yk}^{(1)}-\lambda_{yk}^{(1)-}e^{ik_{\text{TE}}z'}\right]e^{-ik_{\text{TM}}z}.\end{aligned} \quad \text{(A6d)}$$

2) For $D_1 < (z,z') < D_1+D_2$:
   a) $D_1 < z < z'$: with $(i,k)=(x,y,z)$,
   
   $$G_{ik}^{(2)}(z-z') = A_{ik}^{(2)} e^{ik_{2z}z} + B_{ik}^{(2)} e^{-ik_{2z}z}. \quad \text{(A6e)}$$
   
   b) $z' < z < D_1+D_2$: with $(i,k)=(x,y,z)$,
   
   $$\begin{aligned}G_{ik}^{(2)}(z-z') &= \left[A_{ik}^{(2)}+\lambda_{ik}^{(2)+}e^{-ik_{2z}z'}\right]e^{ik_{2z}z}\\&+\left[B_{ik}^{(2)}-\lambda_{ik}^{(2)-}e^{ik_{2z}z'}\right]e^{-ik_{2z}z},\end{aligned} \quad \text{(A6f)}$$
   
   $$\begin{aligned}G_{yk}^{(1)}(z-z') &= \left[A_{yk}^{(1)}+\lambda_{yk}^{(1)+}e^{-ik_{\text{TE}}z'}\right]e^{ik_{\text{TM}}z}\\&+\left[B_{yk}^{(1)}-\lambda_{yk}^{(1)-}e^{ik_{\text{TE}}z'}\right]e^{-ik_{\text{TM}}z},\end{aligned} \quad \text{(A6g)}$$

In Eqs. (A6) above, coefficients $A_{ij}^{(\alpha)}$ and $B_{ij}^{(\alpha)}$ of the plane-wave general solutions implicitly depend on the position $z'$ of the point source, and the particular solutions of the inhomogeneous set of differential equations in each of the materials forming the bilayer are given by:

$$\hat{\lambda}^{(1)\pm} = \begin{pmatrix} \mp i\dfrac{k_x^2-k_0^2\varepsilon_1}{2k_{\text{TM}}k_0^2\varepsilon_1} & 0 & -i\dfrac{k_x}{2k_0^2\varepsilon_1}+\dfrac{\varepsilon'}{2k_{\text{TM}}} \\ 0 & \pm\dfrac{i}{2k_{\text{TE}}} & 0 \\ -i\dfrac{k_x}{2k_0^2\varepsilon_1}-\dfrac{\varepsilon'}{2k_{\text{TM}}} & 0 & \mp i\dfrac{k_{\text{TM}}^2-k_0^2\varepsilon_1}{2k_{\text{TM}}k_0^2\varepsilon_1} \end{pmatrix}, \quad \text{(A7a)}$$

$$\hat{\lambda}^{(2)\pm} = \begin{pmatrix} \mp i\dfrac{k_x^2-k_0^2\varepsilon_2}{2k_{2z}k_0^2\varepsilon_2} & 0 & -i\dfrac{k_x}{2k_0^2\varepsilon_2} \\ 0 & \pm\dfrac{i}{2k_{2z}} & 0 \\ -i\dfrac{k_x}{2k_0^2\varepsilon_2} & 0 & \mp i\dfrac{k_{2z}^2-k_0^2\varepsilon_2}{2k_{2z}k_0^2\varepsilon_2} \end{pmatrix}. \quad \text{(A7b)}$$

## B. Boundary conditions for the Green's functions

The third stage of the solving procedure then consists in writing boundary conditions for the Green's functions at each physical interface separating the different materials in the system. These conditions follow from those concerning tangential $E_x$ and $H_y$ (for TM modes) and $E_y$ and $H_x$ (for TE modes) components of the electric and magnetic fields of light. In the magnetic film, Maxwell's equations yield:

$$H_x = \frac{i\partial_z}{k_0} E_y \text{ and } H_y = \frac{ik_0(\varepsilon' k_x + \varepsilon_1 \partial_z)}{k_x^2 - k_0^2 \varepsilon_1} E_x, \quad (A8)$$

which, combined with boundary conditions for the fields and Eq. (23), lead to boundary conditions for the Green's functions defined in Eqs. (A6) and (A7).

Thus at $z = 0$, with $k = (x, y, z)$:

$$\begin{aligned}
H_x &\to \left. \partial_z G_{yk}^{(1)} \right|_{0^+} = \left. \partial_z G_{yk}^{(V)} \right|_{0^-}, \\
H_y &\to \left. \frac{\varepsilon' k_x + \varepsilon_1 \partial_z}{k_{TE}^2} G_{xk}^{(1)} \right|_{0^+} = \left. \frac{\partial_z}{k_{0z}^2} G_{xk}^{(V)} \right|_{0^-}, \quad (A9) \\
E_y &\to \left. G_{yk}^{(1)} \right|_{0^+} = \left. G_{yk}^{(V)} \right|_{0^-}, \\
E_x &\to \left. G_{xk}^{(1)} \right|_{0^+} = \left. G_{xk}^{(V)} \right|_{0^-}.
\end{aligned}$$

Similarly, at $z = D_1$, with $k = (x, y, z)$:

$$\begin{aligned}
H_x &\to \left. \partial_z G_{yk}^{(1)} \right|_{D_1^-} = \left. \partial_z G_{yk}^{(2)} \right|_{D_1^+}, \\
H_y &\to \left. \frac{\varepsilon' k_x + \varepsilon_1 \partial_z}{k_{TE}^2} G_{xk}^{(1)} \right|_{D_1^-} = \left. \frac{\varepsilon_2 \partial_z}{k_{2z}^2} G_{xk}^{(2)} \right|_{D_1^+}, \quad (A10) \\
E_y &\to \left. G_{yk}^{(1)} \right|_{D_1^-} = \left. G_{yk}^{(2)} \right|_{D_1^+}, \\
E_x &\to \left. G_{xk}^{(1)} \right|_{D_1^-} = \left. G_{xk}^{(2)} \right|_{D_1^+}.
\end{aligned}$$

Finally, at $z = D_1 + D_2$, with $k = (x, y, z)$:

$$\begin{aligned}
H_x &\to \left. \partial_z G_{yk}^{(2)} \right|_{(D_1+D_2)^-} = \left. \partial_z G_{yk}^{(V)} \right|_{(D_1+D_2)^+}, \\
H_y &\to \left. \frac{\varepsilon_2 \partial_z}{k_{2z}^2} G_{xk}^{(2)} \right|_{(D_1+D_2)^-} = \left. \frac{\partial_z}{k_{0z}^2} G_{xk}^{(V)} \right|_{(D_1+D_2)^+}, \quad (A11) \\
E_y &\to \left. G_{yk}^{(2)} \right|_{(D_1+D_2)^-} = \left. G_{yk}^{(V)} \right|_{(D_1+D_2)^+}, \\
E_x &\to \left. G_{xk}^{(2)} \right|_{(D_1+D_2)^-} = \left. G_{xk}^{(V)} \right|_{(D_1+D_2)^+}.
\end{aligned}$$

where $k_{0z}$ is the z-component of the wavevector in vacuum.

The Green's functions denoted $G_{ij}^{(V)}$ in Eqs. (A9) and (A11) relate to the strain-dependent contribution to the (reflected or transmitted) electromagnetic fields in the vacuum, and they write:

1) For $z < 0$:

$$G_{ij}^{(V)}(z - z') = B_{ij}^{(V)} e^{-ik_{0z}z}, (i, j) = (x, y, z). \quad (A12a)$$

2) For $z > D_1 + D_2$:

$$G_{ij}^{(V)}(z - z') = A_{ij}^{(V)} e^{ik_{0z}z}, (i, j) = (x, y, z). \quad (A12b)$$

Substituting Eqs. (A6), (A7) and (A12) into Eqs. (A9)–(A11) then leads to the full expressions of the Green's functions in the entire space. These expressions must be separately established for TE and TM modes, as well as for point source positions either in the magnetic film or in the non-magnetic substrate.

Funding sources and acknowledgments. This research is supported partly by the European Union's Horizon 2020 research and innovation programme (Marie Skłodowska-Curie grant No 64434 "MagIC") and the MPNS COST Action MP1403 "Nanoscale Quantum Optics" (Yu.S.D., N.N.D., and I.L.L), and also is supported by a grant (14.Z50.31.0015) from the Ministry of Education and Science of the Russian Federation (Yu.S.D., N.N.D.). Three of the authors (Yu.S.D., N.N.D., I.L.L.) are grateful to École Nationale d'Ingénieurs de Brest for its financial support (14MOESMOMO14).